%% file: jscc4control_cdc2016.tex
\documentclass[letterpaper, 10 pt, conference]{ieeeconf} 
\IEEEoverridecommandlockouts
\input{jscc4control_cdc2016_defs}
\usepackage{graphicx}
\usepackage{ifthen}

\begin{document}

\input{title}

\maketitle


\begin{abstract}
    \input{abstract.tex}
\end{abstract}

\vspace{.2\baselineskip}

\allowdisplaybreaks

\section{Introduction}
\label{s:intro}

\input{intro.tex}


\section{Problem Setup}
\label{s:model}

\input{model.tex}


\section{Low-Delay Joint Source--Channel Coding}
\label{s:JSCC}

\input{jscc.tex}


\section{Control via Low-Delay JSCC}
\label{ss:LTI}

\input{lqg.tex}


\section{Discussion and Future Research}
\label{s:summary}

\input{summary.tex}


\input{jscc4control_cdc2016.bbl}

\end{document}

%% file: jscc4control_cdc2016_defs.tex
\usepackage[tbtags]{amsmath} 
\usepackage{amssymb}  
\usepackage{enumerate}
\usepackage{amsfonts}
\usepackage[caption=false]{subfig}
\usepackage{color}
\usepackage{cite}

\usepackage{hyperref}

\usepackage{epsfig, graphicx, psfrag}

\usepackage{wasysym}

\usepackage{mathtools}
\mathtoolsset{showonlyrefs=true}

\usepackage[normalem]{ulem}
\usepackage{cancel}

\let\oldsqrt\sqrt
\def\sqrt{\mathpalette\DHLhksqrt}
\def\DHLhksqrt#1#2{%
\setbox0=\hbox{$#1\oldsqrt{#2\,}$}\dimen0=\ht0
\advance\dimen0-0.2\ht0
\setbox2=\hbox{\vrule height\ht0 depth -\dimen0}%
{\box0\lower0.4pt\box2}}

\newtheorem{thm}{Theorem}
\newtheorem{corol}{Corollary}
\newtheorem{remark}{Remark}
\newtheorem{scheme}{Scheme}

\providecommand{\thmref}[1]{Theorem~\ref{#1}}

\providecommand{\secref}[1]{Section~\ref{#1}}

\providecommand{\figref}[1]{Fig.~\ref{#1}}

\providecommand{\schemeref}[1]{Scheme~\ref{#1}}

\newcommand{\ie}{i.e.}
\newcommand{\eg}{e.g.}

\DeclareMathOperator{\sign}{sign}

\newcommand{\bm}[1]{\mbox{\boldmath{$#1$}}}

\newcommand{\SNR}{\mathrm{SNR}}

\newcommand{\SDR}{\mathrm{SDR}}

\newcommand{\eff}{{\textrm{eff}}}

\newcommand{\mD}{\mathcal{D}}
\newcommand{\mE}{\mathcal{E}}

\newcommand{\Comment}[1]{}
\newcommand{\old}[1]{}
\newcommand{\rem}[1]{}

\newcommand{\hx}{\hat{x}}
\newcommand{\hs}{\hat{s}}

\newcommand{\ty}{\tilde y}

\newcommand{\bx}{{\bm x}}

\providecommand{\tx}{\tilde{x}}

\newcommand{\Norm}[1]{\left\| #1 \right\|}

\providecommand{\comment}[1]{}

\providecommand{\norm}[1]{\Norm{#1}}

\newcommand{\beqn}[1]{\begin{eqnarray}\label{#1}}
\newcommand{\eeqn}{\end{eqnarray}}
\newcommand{\beq}[1]{\begin{equation}\label{#1}}
\newcommand{\eeq}{\end{equation}}

\renewcommand{\dagger}{\mathit{T}}

\makeatletter
\newcommand{\vast}{\bBigg@{4}}
\newcommand{\Vast}{\bBigg@{4.9}}
\makeatother

\providecommand{\bu}{{\bm u}}

\providecommand{\oLQGcost}{\bar{J}}
\providecommand{\CostLastX}{F}
\providecommand{\CostXs}{Q}
\providecommand{\CostUs}{R}
\providecommand{\E}[1]{\mathbb{E}\left[ #1 \right]}
\providecommand{\us}{\bar{s}}
\providecommand{\hus}{\hat{\us}}
\providecommand{\hus}{\hat{\us}}

\providecommand{\Ks}{{K_S}}
\providecommand{\Kc}{{K_C}}
\providecommand{\nBW}{\Kc}

\providecommand{\hxENC}{\hx^\mathrm{t}}
\providecommand{\hxDEC}{\hx^\mathrm{r}}
\providecommand{\txENC}{\tx^\mathrm{t}}
\providecommand{\txDEC}{\tx^\mathrm{r}}

\providecommand{\Penc}{P^\mathrm{t}}
\providecommand{\Pdec}{P^\mathrm{r}}
\providecommand{\Penctt}{\bar{P}^\mathrm{t}}

\providecommand{\Kenc}{K^\mathrm{t}}

\providecommand{\nEFF}{n^\mathrm{eff}}

\providecommand{\Jdec}{\bar{J}^\mathrm{r}}
\providecommand{\Jenc}{\bar{J}^\mathrm{t}}

\providecommand{\OPTA}{\mathrm{OPTA}}
\providecommand{\SDRopta}{\SDR_\OPTA}

\renewcommand{\Delta}{\omega}

%% file: title.tex
\title{\LARGE \bf Multi-Rate Control over AWGN Channels via \\ Analog Joint Source--Channel Coding}

\author{Anatoly Khina, Gustav M.\ Pettersson, Victoria Kostina and Babak Hassibi 
    \thanks{A.~Khina, V.~Kostina and B.~Hassibi are with the Dept.\ of Electrical Engineering, California Institute of Technology, Pasadena, CA~91125, USA (E-mails: \mbox{{\em \{khina, vkostina, hassibi\}@caltech.edu}}).
    }
    \thanks{
    G.~M.~Pettersson is with the Dept.\ of Aeronautical and Vehicle Engineering, KTH Royal Institute of Technology, SE-100 44, Stockholm, Sweden (E-mail: {\em gupet@kth.se}).
    }
    \thanks{
    The work of A.~Khina was supported in part by a Fulbright fellowship, Rothschild fellowship and has received funding from 
    the European Union's Horizon 2020 research and innovation programme under the Marie Sk\l lodowska-Curie grant agreement No 708932.
    }
    \thanks{
    The work of G.~M.~Pettersson at Caltech was supported by 
    The Boeing Company under the SURF program.
    }
    \thanks{
    The work of B.~Hassibi was supported in part by the National Science Foundation under grants CNS-0932428, CCF-1018927, CCF-1423663 and CCF-1409204, by a grant from Qualcomm Inc., by NASA's Jet Propulsion Laboratory through the President and Director’s Fund, and by King Abdullah University of Science and Technology.    
    }
} 

%% file: abstract.tex
We consider the problem of controlling an unstable plant over an additive white Gaussian noise (AWGN) channel with a transmit power constraint, where the signaling rate of communication is larger than the sampling rate (for generating observations and applying control inputs) of the underlying plant. Such a situation is quite common since sampling is done at a rate that captures the dynamics of the plant and which is often much lower than the rate that can be communicated. This setting offers the opportunity of improving the system performance by employing multiple channel uses to convey a single message (output plant observation or control input). Common ways of doing so are through either repeating the message, or by quantizing it to a number of bits and then transmitting a channel coded version of the bits whose length is commensurate with the number of channel uses per sampled message. We argue that such ``separated source and channel coding" can be suboptimal and propose to perform joint source--channel coding. Since the block length is short we obviate the need to go to the digital domain altogether and instead consider analog joint source--channel coding. For the case where the communication signaling rate is twice the sampling rate, we employ the Archimedean bi-spiral-based
Shannon--Kotel'nikov analog maps to show significant improvement in stability margins and linear-quadratic Gaussian (LQG) costs over simple schemes that employ repetition.

%% file: intro.tex
Networked control systems, especially those for which the links connecting the different components of the system 
(plant, observer, and controller, say) are noisy, 
are increasingly finding applications and, as a result have been the subject of intense recent investigations \cite{NetworkedControlSurvey_ProcIEEE,GuptaDanaHespanhaMurrayHassibi_EstimationControl,SchenatoSinopoliFranceschettiPoolaSSS}. 
In many of these applications the rate at which the output of the plant is sampled and observed, 
as well as the rate at which control inputs are applied to the plant, 
is different from the signaling rate with which communication occurs. 
We shall henceforth call such systems {\em multi-rate} networked control systems. 
The rate at which the plant is sampled and controlled is often governed by how fast the dynamics of the plant is, 
whereas the signaling rate of the communication depends on the bandwidth available, the noise levels, etc. 
As a result, there is no inherent reason why these two rates should be related and, 
in fact, the communication rate is almost always higher than the sampling rate. 

This latest fact clearly gives us the opportunity to improve the performance of the system 
by having the possibility to convey information about each sampled output of the plant, 
and/or each control signal, through {\em multiple} uses of the communication channel. 
An obvious strategy is to simply repeat the transmitted signal (so-called {\em repetition coding}). 
In analog communication this simply adds a linear factor to the SNR (3 dB for a single repetition); 
in digital communication over a memoryless packet erasure link, say, 
it simply reduces the probability of packet loss exponentially in the number of retransmissions. 
A more sophisticated solution would be to first quantize the analog message 
(the sampled output or the control signal) 
and then protect the quantized bits with an error-correcting channel code whose block length is commensurate 
with the number of channel uses available per sample. 
A yet more sophisticated solution would be to use a {\em tree code} which collectively encodes the quantized bits in a causal fashion over all channel uses \cite{SahaiMitterPartI,SukhavasiHassibi,TreeCodes:ISIT2016}. 

The latter two solutions implicitly assume what is called the ``separation between source and channel coding", 
i.e., that quantization of the messages and channel coding of the quantized bits (using either a block code or a tree code) 
can be done independently of one another. While this is asymptotically true in communication systems 
(where it is a celebrated result), 
it is not true for control systems where the overall objective is to minimize a linear-quadratic Gaussian (LQG) cost~\cite{TatikondaSahaiMitter}. 
To minimize an LQG cost what is needed is joint source--channel coding (JSCC). 
Unfortunately, in its full generality, 
this is known to be a notoriously difficult problem and so it has rarely been attempted (especially, in a control context).
Nonetheless, this is what we shall attempt in this paper. 

We assume the communication links are AWGN (additive white Gaussian noise) channels with a certain signal-to-noise ratio (SNR). 
As we show below, this SNR puts an upper limit on the size of the maximum unstable eigenvalue of the plant that can be stabilized.
We further assume that the signaling rate of the communication channel is not much larger than the sampling rate of the plant, 
say only a factor of 2 to 10 larger. 
Thus, if one sets aside the (daunting) task of performing coding over multiple messages (\emph{a la} tree codes) 
then one is left with constructing a joint source--channel code of relatively short length~--- something that could very well be feasible. 
In particular, since both the message and transmitted signals are analog, 
in this short block regime it is not even clear whether it is necessary to go through a digitization process. 
Thus, we shall focus on {\em analog} JSCC, originally proposed by Shannon~\cite{Shannon49} and Kotel'nikov~\cite{KotelnikovJSCC}, 
which can simply be viewed as an appropriately chosen nonlinear mapping from the analog message to the analog transmitted signal(s). 

Finally, we should mention that we view this work as a first step and the results as preliminary. 
Nonetheless, these already indicate that one can obtain substantial gains (in the LQG cost) over simple schemes, 
such as repetition, by using the ideas mentioned above. 
The design of more sophisticated JSCC schemes, as well as a comprehensive comparison of different schemes will be deferred to future work.

%% file: model.tex
\begin{figure}[t]
\centering
\vspace{-7\baselineskip}
    \includegraphics[width=\columnwidth]{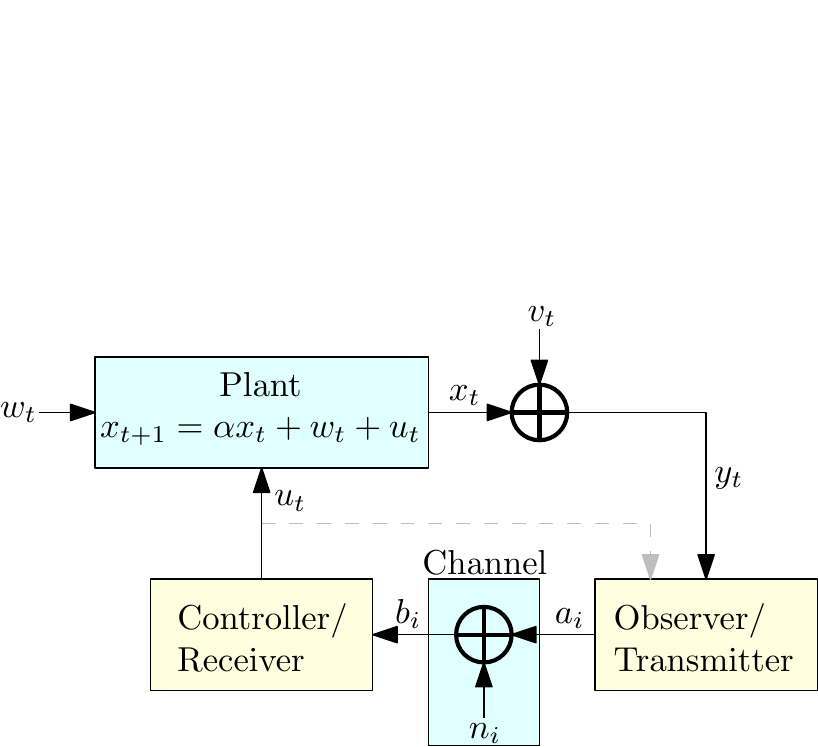}
\centering
    \caption{Scalar control system with white driving and observation AWGNs and an AWGN communication channel. 
	     The dashed line represents the assumption that the past control signals are available at the transmitter.}
    \label{fig:model}
\end{figure}

We now formulate the control--communication setting that will be treated in this work, depicted also in \figref{fig:model}.
We concentrate on the simple case of a scalar full observable state and a scalar AWGN channel.
The model and solutions can be extended to more complex cases of vector states and 
multi-antenna channels.

Consider the scalar system with the plant evolution:
\begin{align}
\label{eq:plant}
    x_{t+1} = \alpha x_t + w_t + u_t ,
\end{align}
where $x_t$ is the (scalar) state at time $t$, 
$w_t$ is an AWGN of power $W$, $\alpha > 1$ is a known scalar, 
and $u_t$ is the control signal. Assume further that $x_0$ is Gaussian with power $P_0$.

The measured output is equal to the state corrupted by noise:
\begin{align}
\label{eq:observer}
    y_t = x_t + v_t,
\end{align}
where $v_t$ is an AWGN of power $V$.

In contrast to classical control settings, 
the observer and the controller are not co-located, and are connected instead via an AWGN channel
\begin{align}
\label{eq:channel}
    b_i = a_i + n_i ,
\end{align}
where $b_i$ is the channel output, $a_i$ is the channel input subject to a unit power constraint, 
and $n_i$ is an AWGN of power $1/\SNR$.\footnote{This representation is w.l.o.g., as the case of an average power $P_C$ and 
noise power $N$, can always be transformed to an equivalent channel with average power $1$ and noise power $N/P_C \triangleq 1/\SNR$
by multiplying both sides of \eqref{eq:channel} by $1/\sqrt{P_C}$.}

We assume an integer ratio $\nBW$ between the sample rate of~\eqref{eq:observer} and the signaling rate over the channel \eqref{eq:channel}. 
That is, $\nBW$ channel uses \eqref{eq:channel} are available per one control sample~\eqref{eq:plant}, \eqref{eq:observer}.

In this work we further assume that the observer knows all past control signals $\{u_i | i = 1, \ldots, t-1\}$;
for a discussion of the case when such information is not available at the observer, see \secref{s:summary}.

Similarly to the classical LQG control (in which the controller and the observer are co-locatedco-located controller and observer), we wish to minimize the average stage LQG cost after the total number of observed samples $T$:
\begin{align}
\label{eq:LQG:cost}
    \oLQGcost_T \triangleq \frac{1}{T} \E{ \CostLastX x_{T+1}^2 + \sum_{t=1}^T \Big( \CostXs x_t^2 + \CostUs u_t^2 \Big) } ,
\end{align}
for some non-negative constants $F$, $Q$ and $R$, 
by designing appropriate operations at the observer [which also plays the role of the transmitter over \eqref{eq:channel}] and the controller [which also serves as the receiver of \eqref{eq:channel}].
The infinite horizon cost is defined as 
\begin{align}
    \oLQGcost_\infty \triangleq \lim_{T \to \infty} \oLQGcost_T \,.
\end{align}

To that end, we recall next known results from information theory for joint source--channel coding design with low delay.

%% file: jscc.tex
In this section, we review known results from information theory and communications for transmitting an i.i.d.\ 
zero-mean Gaussian source $s_t$ of power $P_S$ over an AWGN channel~\eqref{eq:channel}.

The number of source samples generated per a time instant is not necessarily equal to the channel uses of \eqref{eq:channel} per the same time. In general, consider the case where $\Kc$ channel uses of $b_i$ are available for every $K_S$ source samples of $s_t$.

The goal of the transmitter is to convey the source $s_t$ to the receiver with a minimal possible average distortion, 
where the appropriate distortion measure for our case of interest is the mean square error distortion.

To that end, the transmitter applies a mapping $\mE$ that transforms every $K_S$ source samples to $\Kc$
channel inputs:
\begin{align}
    \left( a_1, \ldots, a_\Kc \right) = \mE\left( s_1, \ldots, s_\Ks \right) ,
\end{align}
such that the input power constraint is satisfied:
\begin{align}
    \frac{1}\Ks \E{ \left\{ \mE\left( s_1, \ldots, s_\Ks \right) \right\}^2 } \leq 1 .
\end{align}
The receiver, upon receiving the $\Kc$ channel outputs of \eqref{eq:channel}~--- corresponding to the $\Kc$ transmitted channel inputs~--- applies a mapping to these measured outputs to recover estimates $\hs_t$ of the source samples:
\begin{align}
    \left( \hs_1, \ldots, \hs_\Ks \right) = \mD \left( b_1, \ldots, b_\Kc \right) .
\end{align}

The resulting average distortion of this scheme is 
\begin{align}
\label{eq:D}
    D = \frac{1}\Ks \sum_{t=1}^\Ks \E{ \left( s_t - \hs_t \right)^2 } , 
\end{align}
and the corresponding (source) signal-to-distortion ratio (SDR) is defined as 
\begin{align}
    \SDR \triangleq \frac{P_S}{D} \,.
\end{align}

Our results here are more easily presented in terms of unbiased errors, as these can be regarded 
as uncorrelated additive noise in the sequel (when used as part of the developed control scheme).
Therefore, we consider the use of (sample-wise) correlation-sense unbiased estimators (CUBE), namely, 
estimators that satisfy 
\begin{align}
    \E{s_t \left( s_t - \hs_t \right)} = 0 .
\end{align}
We note that any estimator $\hs^B_t$
can be transformed into a CUBE $\hs_t$ by multiplying by a suitable constant:
\begin{align}
    \hs_t = \frac{ \E{s_t^2} }{ \E{s_t \hs^B_t} } \hs^B_t ;
\end{align}
for a further discussion of such estimators and their use in communications the reader is referred to~\cite{RematchAndForward_Full}.

Shannon's celebrated result~\cite{Shannon48} states that the minimal achievable distortion, using \emph{any} transmiter--receiver scheme, 
is dictated, in the case of a Gaussian source, by\footnote{The rate--distortion function here is written in terms of the unbiased SDR, in contrast to the more common biased SDR expression $\log(\SDR)$.}
\begin{align}
\label{eq:separation}
\!\!\!
    \tfrac{\Ks}{2} \log\left( 1 + \SDR \right) = \Ks R(D) \leq \Kc C = \tfrac{\Kc}{2} \log\left( 1 + \SNR \right)
\end{align}
where $R(D)$ is the rate--distortion function of the source and $C$ is the channel capacity~\cite{Shannon48}.
Thus, the optimal SDR, commonly referred to as \emph{optimum performance theoretically achievable} (OPTA) SDR, 
is given by
\begin{align}
\label{eq:SDR:OPTA}
    \SDRopta = \left( 1 + \SNR \right)^{\Kc / \Ks} - 1 .
\end{align}
Shannon's proof (for a more general case of not necessarily Gaussian source or channel) is based upon the \emph{separation principle}, according to which the source samples are partitioned into blocks and quantized together, resulting in (approximately) uniform independent bits. These bits are then partitioned again into blocks and encoded together to form the channel inputs. At the receiver first the coded bits are recovered, followed by the reconstruction 
of the source samples from these bits. 

However, this compression--coding separation-based technique is optimal only in the limit when the blocklengths $\Ks$ and $\Kc$  grow to infinity for a fixed ratio between the two, which implies, in turn, very large delays.

For finite blocklengths, 
\eqref{eq:SDR:OPTA}
cannot be exactly attained, 
except for specific cases in which the source and the distortion measure are probabilistically matched to the channel~\cite{GastparToCodeOrNot}, and strictly tighter outer bounds on the distortion can be derived \cite{ZivZakaiJSCC,IngberLeibowitzISIT2008,TridenskiZamirIngber}.
One eminent case where such a matching occurs is that of a Gaussian source and a Gaussian channel with matching number of samples/uses $\Kc = \Ks$.
In this case, sending each source sample as is, up to a possible power adjustment, 
proves optimal and achieves~\eqref{eq:SDR:OPTA}
with $\Kc = \Ks = 1$ (and hence also any other positive integer)~\cite{Goblick65}.
Unfortunately, this breaks down when $\Kc \neq \Ks$, and consequently led to the proposal and study of various techniques 
for low-delay JSCC schemes.\footnote{The term JSCC is somewhat misleading, as in many of these schemes there is no use of digital components, let alone coding, including the Shannon--Kotel'nikov (SK) maps which are described in detail and used in the sequel.}

We next concentrate on the simple case of $\Ks = 1$ and $\Kc = 2$. 
That is, the case in which one source sample is conveyed over two channel uses.

A na\"ive approach is to send the source as is over both channel uses, up to a power adjustment.
The corresponding unbiased SDR in this case is 
\begin{align}
\label{eq:SDR:linear}
    \SDR_\text{lin} = 2 \SNR , 
\end{align}
a linear improvement rather than an exponential one as in \eqref{eq:SDR:OPTA}.
This scheme approaches \eqref{eq:SDR:OPTA} for very low SNRs, but suffers great losses at high SNRs.
We note that the linear factor 2 comes from the fact that the total power available over both channel uses has doubled, 
and the same performance can be attained by allocating all of the available power to the first channel use and remaining silent during the second channel use. 

\begin{figure}[t]
\centering
\includegraphics[width=\columnwidth]{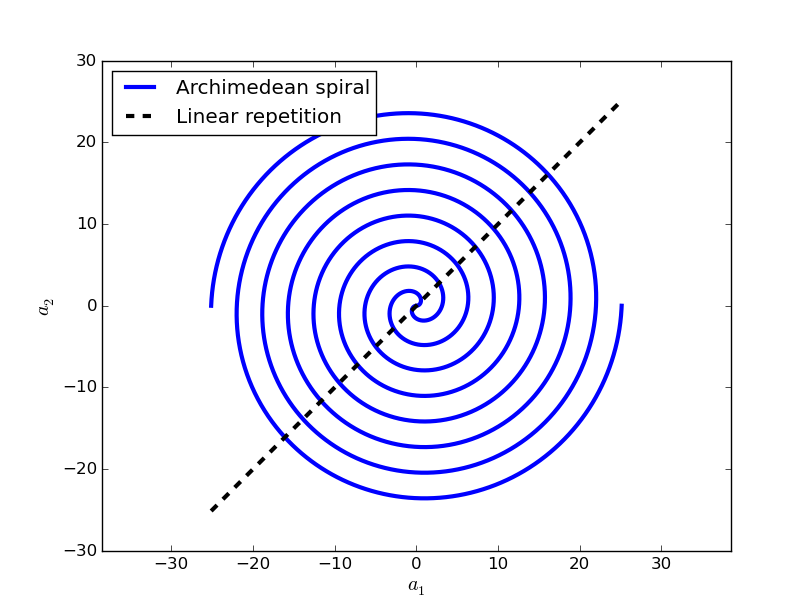}
\centering
    \caption{Linear repetition and Archimedean spiral curves.}
    \label{fig:spiral}
\end{figure}

\begin{figure}[t]
    \centering
    \subfloat[$\lambda = 1$.]
    {
    \label{fig:spiral:performance:lambda=1}
	\includegraphics[width=\columnwidth]{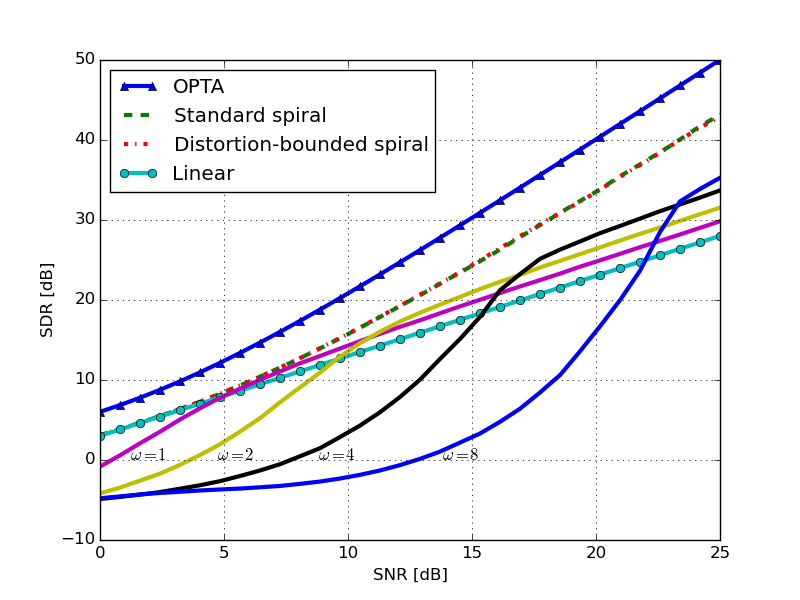}
    }
    \\
    \subfloat[$\lambda = 0.5$.]
    {
    \label{fig:spiral:performance:lambda=0.5}
	\includegraphics[width=\columnwidth]{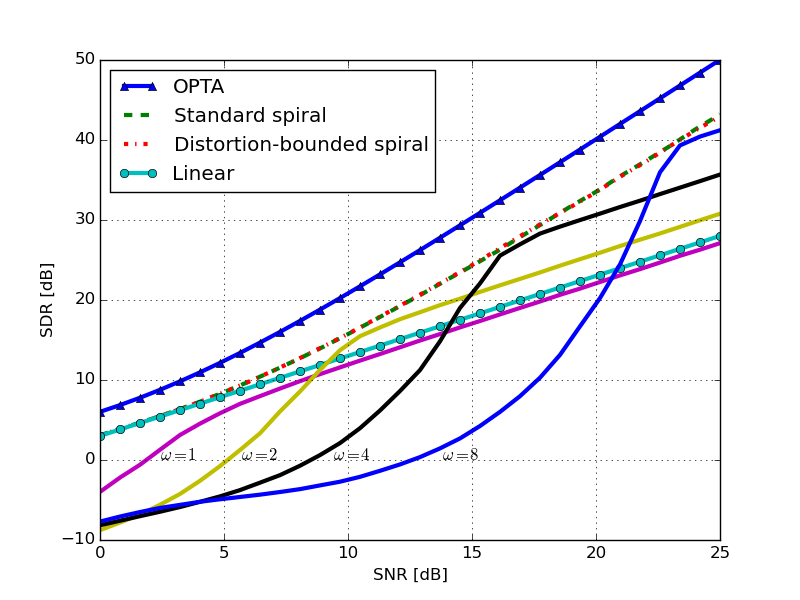}
    }
    \caption{Performances of the JSCC linear repetition scheme, OPTA bound, and the JSCC SK spiral scheme
    for optimized $\lambda$ and $\Delta$, for the standard case ($\beta = 1$) and distortion-bounded case.
    The solid lines depict the performance of the standard spiral for various values of $\Delta$
    for two stretch parameters $\lambda = 0.5$ and $1$, which perform better at high and low SNRs,
     respectively.}
\label{fig:spiral:performance}
\end{figure}

This suggests that better mappings that truly exploit the extra channel use can be constructed.
The first to propose an improvement for the 1:2 case were Shannon~\cite{Shannon49} and Kotel'nikov~\cite{KotelnikovJSCC}, in the late 1940s.
In their works, the source sample is viewed as a point on a single-dimensional line, 
whereas the two channel uses correspond to a two-dimensional space. 
In these terms, the linear scheme corresponds to mapping the one-dimensional source line to a 
straight line in the two-dimensional channel space (see~\figref{fig:spiral}), and hence clearly cannot provide any improvement 
(since AWGN is invariant to rotations). 
However, by mapping the one-dimensional source line into a two-dimensional curve that fills better the space, a great boost in performance can be attained. 
Specifically, consider the Archimedean bi-spiral, which was considered in several works~\cite{ChungPhD,ShannonKotelnikovMaps_Ramstad,ShannonKotelnikovMaps_MMSE_Dec,KvecherRephaeliJSCC}
(depicted in \figref{fig:spiral}):
\begin{align}
\label{eq:spiral:reg}
    \begin{cases}
	a^\text{reg}_1(s) =  c^\text{reg}\, s \, \cos( \Delta s) \phantom{\sign(s)}
	= c^\text{reg}\, |s| \, \cos( \Delta |s|) \sign(s)
     \\ a^\text{reg}_2(s) =  c^\text{reg}\, s \, \sin( \Delta s) \sign(s)
     =  c^\text{reg}\, |s| \, \sin( \Delta |s|) \sign(s)
    \end{cases}
\end{align}
where $\Delta$ determines the rotation frequency, the factor $c^\text{reg}$ is chosen to satisfy the power constraint,
and the $\sign(s)$ term is needed 
to avoid overlap of the curve for positive and negative values of $s$ (for each of which now corresponds a distinct spiral, and the two meet only at the origin).
This spiral allows to effectively improve the resolution w.r.t.\ small noise values, since the one-dimensional 
source line is effectively stretched compared to the noise, 
and hence the noise magnitude shrinks when the source curve is mapped (contracted) back. 
However, for large noise values, a jump to a different branch~--- referred to as a \emph{threshold effect}~--- may occur, incurring a large distortion.
Thus, the value $\Delta$ needs to be chosen to be as large as possible to allow maximal stretching of the curve for the same given power, while maintaining a low threshold event probability.
The SDRs for different values of $\Delta$ are depicted in \figref{fig:spiral:performance:lambda=1}.

Another ingredient that is used in conjunction with \eqref{eq:spiral:reg} is stretching $s$ prior to mapping it to a bi-spiral using \mbox{$\phi_\lambda(s) \triangleq \sign(s) |s|^\lambda$}: 
\begin{align}
\label{eq:spiral:stretch}
    \!\!\!\!
    \begin{cases}
	a_1^\text{stretch}(s) = a^\text{reg}_1(\phi_\lambda(s)) = c^\text{stretch} |s|^\lambda \cos \left( \Delta |s|^\lambda \right) \sign(s)
     \\ a_2^\text{stretch}(s) = a^\text{reg}_2(\phi_\lambda(s)) = c^\text{stretch} |s|^\lambda \sin \left( \Delta |s|^\lambda \right)  \sign(s)
     \end{cases}
\end{align} 

The choice $\lambda = 0.5$ promises great boost in performance in the region of high SNRs, as is seen in \figref{fig:spiral:performance:lambda=0.5}.
We further note that although the optimal decoder is a minimum mean square error (MMSE) estimator $\E{s | b_1, b_2}$, in this case, 
the maximum-likelihood (ML) decoder, $p(b_1, b_2|s)$, 
achieves similar performance for moderate and high SNRs.
A joint optimization of $\lambda$ and $\Delta$ for each SNR, for both ML and MMSE decoding, was carried out in~\cite{ShannonKotelnikovMaps_MMSE_Dec} and is depicted in \figref{fig:spiral:performance}.

A desired property of the linear JSCC schemes is their SDR proportional improvement with the channel SNR (``SNR universality'').
Such an improvement is not allowed by the separation-based technique, 
as it fails when the actual SNR is lower than the design SNR, 
and does not promise any improvement for SNRs above it.
This motivated much work in designing JSCC schemes whose performance improves with the SNR, 
even for the case of large blocklengths~\cite{MittalPhamdo,Reznic,AnalogMatching}.
The schemes in these works achieve optimal performance \eqref{eq:SDR:OPTA} for a specific design SNR~\eqref{eq:SDR:OPTA}, 
and improve linearly for higher SNRs. 
Similar behavior is observed also in \figref{fig:spiral:performance} where the optimal $\Delta$ value 
varies with the (design) SNR, and mimics closely the quadratic growth in the SDR.
Above the design SNR, linear growth is achieved for a particular choice of $\Delta$.

We further note that the distortion component incurred when a threshold event happens, 
grows with $|s|$. To avoid this behavior, instead of increasing the magnitude 
$\norm{a^\text{stretch}}$
proportionally to the phase $\angle \left( a^\text{stretch} \right)$ as in \eqref{eq:spiral:stretch}, we increase it slightly faster at a pace that guarantees that the incurred distortion does not grow with $|s|$:
\begin{align}
\label{eq:spiral:bounded}
    \begin{cases}
	a_1^\text{bounded}(s) = c^\text{bounded} |s|^{\lambda \beta} \cos \left( \Delta |s|^\lambda \right) \sign(s)
     \\ a_2^\text{bounded}(s) = c^\text{bounded} |s|^{\lambda \beta} \sin \left( \Delta |s|^\lambda \right) \sign(s)
     \end{cases}
     \\[-1.\baselineskip]
\end{align} 
for some $\beta > 1$. This has only a slight effect on the resulting SDRs, as is illustrated in \figref{fig:spiral:performance}.

Finally note that in no way do we claim that the spiral-based Shannon--Kotel'nikov (SK) scheme is optimal. 
Various other techniques exist, most using a hybrid of digital and analog components~\cite{KleinerRimoldiGLOBECOM2009,Tuncel_ZeroDelayJSCCwithWZ,AkyolViswanathaRoseRamstad_IterativeJSCC}, 
which outperform the spiral-based scheme for various parameters. Nevertheless, this scheme 
is the earliest technique to be considered and gives good performance boosts which suffice for our demonstration.

%% file: lqg.tex
In this section we construct a Kalman-filter-like solution~\cite{BertsekasControlVol1} by employing JSCC schemes.
We note that the additional complication here is due to the communication 
channel \eqref{eq:channel} 
and its inherent input power constraint.

Denote by $\hxDEC_{t_1|t_2}$ the estimate of $x_{t_1}$ at the receiver given $\{b_i | i = 1, \ldots, t_2\}$, where `r' stands for `receiver', and by $\hxENC_{t_1|t_2}$ the estimate of $x_{t_1}$ 
given $\{y_i | i = 1, \ldots, t_2\}$, where `t' stands for `transmitter'. Denote further 
their mean square errors (MSEs) by $\Pdec_{t_1|t_2} \triangleq \E{\txDEC_{t_1|t_2}}$ and $\Penc_{t_1|t_2} \triangleq \E{\txENC_{t_1|t_2}}$, where $\txENC_{t_1|t_2} \triangleq x_{t_1} - \hxENC_{t_1|t_2}$
and $\txDEC_{t_1|t_2} \triangleq x_{t_1} - \hxDEC_{t_1|t_2}$.

Then, the scheme works as follows. At time instant $t$, the controller constructs an estimate $\hxDEC_{t|t}$ of $x_t$. It then applies the control signal $u_t = -L_t \hxDEC_{t|t}$ to the plant, 
for a pre-determined gain $L_t \neq 0$.
Note that, since both the controller and the observer know the previously applied control signals $\{u_j | j = 1, \ldots, t \}$, 
they also know $\hxDEC_{t|t}$ and $\hxDEC_{t+1|t}$.

Hence, in order to describe $x_t$, the controller aims to convey its best estimate of the state $\hxENC_{t|t}$.
To that end, it can save transmit power by transmitting the error signal $(\hxENC_{t|t} - \hxDEC_{t|t-1})$, instead of $\hxENC_{t|t}$.
The controller can then add back $\hxDEC_{t|t-1}$ to the received signal to construct $\hxDEC_{t|t}$.

\begin{remark}
    Note that even in the case of a fully observable state, \ie, when $V = 0$, 
    the state is corrupted by $n_t$ when conveyed over the AWGN channel \eqref{eq:channel} to the controller. 
    The performance of the transmission and the estimation processes applied by the observer and the controller, respectively, determine in turn, the total effective \emph{observation noise}.
\end{remark}

The general scheme used throughout this work is detailed below.

\begin{scheme}
\label{scheme:general}
\ \\

\vspace{-.5\baselineskip}
    \textbf{Observer/Transmitter:} At time $t$
    \begin{itemize}
    \item
	Generates the desired error signal 
	\begin{subequations}
	\label{eq:LQG:source}
	\noeqref{eq:LQG:source:diff,eq:LQG:source:explicit,eq:LQG:source:diff:eff}
	\begin{align}
	    s_t &= \hxENC_{t|t} - \hxDEC_{t|t-1}
	\label{eq:LQG:source:diff}
	 \\ &= \txDEC_{t|t-1} - \txENC_{t|t}
	\label{eq:LQG:source:diff:eff}
	\end{align}
	\end{subequations}
	of average power $\Pdec_{t|t-1} - \Penc_{t|t}$ (determined in the sequel).
    \item
	Since the channel input is subject to a unit power constraint, $s_t$ is normalized:
	\begin{align}
	\label{eq:LQG:source:normalized}
	    \us_t = \frac{1}{\sqrt{\Pdec_{t|t-1} - \Penc_{t|t}}} s_t \,.
	    \\[-1.4\baselineskip]
	\end{align}
    \item
	Constructs $\Kc$ channel inputs $a_i$ corresponding to $\us_t$, 
	using a bounded-distortion JSCC scheme of choice of rate ratio $1 : \Kc$ 
	with (maximum given any input) average distortion $1 / \SDR_0$ for the given channel SNR.
    \item
	Sends the $\Kc$ channel inputs $a_i$ over the channel \eqref{eq:channel}.
    \end{itemize}

\vspace{.5\baselineskip}
    \textbf{Controller/Receiver:} At time $t$ 
    \begin{itemize}
    \item
	Receives the $\Kc$ channel outputs $b_i$ corresponding to time sample $t$.
    \item
	Recovers a CUBE of the source signal $\us_t$: $\hus_t = \us_t + \nEFF_t$, 
	where $\nEFF_t \perp \us_t$ is an additive noise of power of (at most) $1 / \SDR_0$.
    \item 
	Unnormalizes $\hus_t$ to construct an estimate of $s_t$: 
	\begin{subequations}
	\label{eq:LQG:scheme:effective_channel}
	\noeqref{eq:LQG:scheme:effective_channel:unnormalized,eq:LQG:scheme:effective_channel:explicit,eq:LQG:scheme:effective_channel:substitute,eq:LQG:scheme:effective_channel:AWGN}
	\begin{align}
	    \hs_t &= \sqrt{\Pdec_{t|t-1} - \Penc_{t|t}} \, \hus_t
	\label{eq:LQG:scheme:effective_channel:unnormalized}
	 \\ &= \sqrt{\Pdec_{t|t-1} - \Penc_{t|t}} \left( \us_t + \nEFF_t \right)
	\label{eq:LQG:scheme:effective_channel:substitute}
	 \\ &= \hxENC_{t|t} - \hxDEC_{t|t-1} + \sqrt{\Pdec_{t|t-1} - \Penc_{t|t}} \, \nEFF_t .
	\label{eq:LQG:scheme:effective_channel:explicit}
	\end{align}
	\end{subequations}
	
    \item
	Constructs an estimate $\hxDEC_{t|t}$ of $x_t$ from all received channel outputs until and including at time $t$. Since $\hs_t \perp \hxDEC_{t|t-1}$, the linear MMSE estimate amounts to\footnote{If the resulting effective noise $n_t^\eff$ is not an AWGN with power that does not depend on the channel input, then a better estimator than that in \eqref{eq:scheme:DEC:estimate} may be constructed.}
	\begin{align}
	\label{eq:scheme:DEC:estimate}
	    \hxDEC_{t|t} &= \hxDEC_{t|t-1} + \frac{\SDR_0}{1 + \SDR_0} \hs_t \,,
	\end{align}
	with an MSE of 
	\begin{align}
	\label{eq:scheme:DEC:estimation:error}
	    \Pdec_{t|t} = \frac{1}{1 + \SDR_0} \left( \Pdec_{t|t-1} + \SDR_0 \, \Penc_{t|t} \right) .
	\end{align}
    \item
	Generates the control signal ($L_t$ is given next): 
	\begin{align}
	    u_t = -L_t \hxDEC_{t|t} \,, 
	\end{align}
	and the receiver prediction of the next system state 
	\begin{align}
	    \hxDEC_{t|t-1} = \alpha \hxDEC_{t-1|t-1} + u_{t-1} \,.
	\end{align}
    \end{itemize}
\end{scheme}

\vspace{.5\baselineskip}
The control (LQG) signal gain $L_t$ is given by (see, \eg, \cite{BertsekasControlVol1}):
\begin{subequations}
\label{eq:LQG:ControlGain}
\noeqref{eq:LQG:ControlGain:L,eq:LQG:ControlGain:S,eq:LQG:ControlGain:S0}
\begin{align}
	L_t &= \frac{\alpha S_{t+1}}{S_{t+1} + R} \,,
\label{eq:LQG:ControlGain:L}
     \\[.5ex] S_t &=  \frac{\alpha^2 R\, S_{t+1}}{S_{t+1} + R} + Q \,,
\label{eq:LQG:ControlGain:S}
     \\[.5ex]
     S_T &= F \,.
\label{eq:LQG:ControlGain:S0}
\end{align}
\end{subequations}
Using \eqref{eq:scheme:DEC:estimation:error} and \eqref{eq:plant}, 
the prediction error at the decoder is given by the following recursion:
\begin{align}
\label{eq:scheme:DEC:prediction:error}
    \Pdec_{t+1|t} = \frac{\alpha^2}{1 + \SDR_0} \left( \Pdec_{t|t-1} + \SDR_0 \, \Penc_{t|t} \right) + W .
\end{align}

The estimates $\hxENC_{t|t}$ can be generated via Kalman filtering (see, \eg, \cite{BertsekasControlVol1}):
\label{eq:scheme:ENC:estimate}
\noeqref{eq:scheme:ENC:estimate:residual,eq:scheme:ENC:estimate:estimator}
\begin{subequations}
\begin{align}
    \ty_t &= y_t - \alpha \hxENC_{t-1|t-1} - u_{t-1} \,,
 \label{eq:scheme:ENC:estimate:residual}
 \\ \hxENC_{t|t} &= \alpha \hxENC_{t-1|t-1} + u_{t-1} + \Kenc_t \ty_t \,,
 \label{eq:scheme:ENC:estimate:estimator}
\end{align}
\end{subequations}
where the Kalman filter coefficients are generated via the recursion~\cite{BertsekasControlVol1}:
\begin{subequations}
\label{eq:scheme:ENC:filters}
\noeqref{eq:scheme:ENC:filters:K,eq:scheme:ENC:filters:P,eq:scheme:ENC:filters:Ptt}
\begin{align}
    \Kenc_t &= \frac{\Penc_{t|t-1}}{\Penc_{t|t-1} + V} \,,
\label{eq:scheme:ENC:filters:K}
 \\ \Penc_{t+1|t} &= \alpha^2 \Penc_{t|t-1} \left( 1 - \Kenc_t \right) + W ,
\label{eq:scheme:ENC:filters:P}
 \\ \Penc_{t|t} &= \Kenc_t V .
\label{eq:scheme:ENC:filters:Ptt}
\end{align}
\end{subequations}

The recursive relations~\eqref{eq:scheme:DEC:prediction:error} and~\eqref{eq:scheme:ENC:filters} lead to the following condition for the stabilizability of the control system.
\begin{thm}[Achievable]
\label{thm:UB}
    The scalar control system of \secref{s:model} is stabilizable using \schemeref{scheme:general} 
    if 
    $\alpha^2 < 1 + \SDR_0$, and its infinite-horizon average stage LQG cost $\Jdec$ is upper bounded by 
    \begin{subequations}
    \label{eq:thm:UB:cost}
    \noeqref{eq:thm:UB:cost:DEC,eq:thm:cost:ENC}
    \begin{align}
        \Jdec &\leq \Jenc + \frac{Q + \left( \alpha^2 - 1 \right) S}{1 + \SDR_0 - \alpha^2} \left( \Penc - \Penctt \right) ,
    \label{eq:thm:UB:cost:DEC}
     \\ \Jenc &= Q \Penctt + S \left( \Penc - \Penctt \right) ,
    \label{eq:thm:cost:ENC}
    \end{align}
    \end{subequations}
    where $\Jenc$ is the average stage cost achievable at the observer,\footnote{Alternatively, this is the cost $\Jdec$ in the limit $\SNR \to \infty$.} 
    and $\Penc$ and $\Penctt$ are the infinite-horizon values of $\Penc_{t|t-1}$ and $\Penc_{t|t}$, respectively; 
    $S$ and 
    $\Penc$ are given as the positive solutions of 
    \begin{align}
	S^2 - \left[ Q + \left( \alpha^2 - 1 \right) R \right]S - QR &= 0
    \end{align}
    and 
    \begin{align}
	\left( \Penc \right)^2 - \left[ \left( \alpha^2 - 1 \right) V + W \right] \Penc - V W &= 0
    \end{align}
    respectively, and 
    \begin{align}
        \Penctt = \frac{\Penc V}{\Penc + V} \,.
    \end{align}
    \vspace{.0001\baselineskip}
\end{thm}

\vspace{-.9\baselineskip}
The following theorem is an adaptation of the lower bound in \cite{KostinaHassibi:Allerton2016} to our setting of interest.

\begin{thm}[Lower bound]
\label{thm:LB}
    The scalar control system of \secref{s:model} is stabilizable 
    only if $\alpha^2 < 1 + \SDRopta$, 
    and the optimal achievable infinite-horizon average stage LQG cost is lower bounded by 
    \begin{align}
    \label{eq:thm:LB:cost}
        \Jdec &\geq \Jenc + \frac{Q + \left( \alpha^2 - 1 \right) S}{1 + \SDRopta - \alpha^2} \left( \Penc - \Penctt \right) ,
    \end{align}
    where $\Penc$, $\Penctt$ and $S$ are as in \thmref{thm:UB}, and $\SDRopta$ is given in 
    \eqref{eq:SDR:OPTA}.
\end{thm}

By comparing \eqref{eq:thm:UB:cost:DEC} and \eqref{eq:thm:LB:cost} we see that the possible gap between the two bounds stems from the gap between the bounds on the achievable SDR over the AWGN channel~\eqref{eq:channel}.

It is interesting to note that in this case, in stark contrast to the classical LQG setting 
in which the system is stabilizable for \emph{any values} of $\alpha$, $V$ and $W$, 
low values of SDR render the system unstable. 
Hence, it provides, among others, the minimal required transmit power for the system to remain stable.
The difference from the classical LQG case stems from the additional input power constraint, 
which effectively couples the power of the observation noise with that of the estimation error, 
and was previously observed in, \eg, \cite{LQGoverAWGN:linear:NoBraslavsky,LQGoverAWGN:linear:WithBraslavsky,KostinaHassibi:Allerton2016,TatikondaSahaiMitter}
for the fully-observed setting.

\begin{remark}
    In the limit $\SNR \to \infty$, we attain also $\SDR \to \infty$. In this case, 
    the estimate~\eqref{eq:scheme:DEC:estimate} and the prediction error~\eqref{eq:scheme:DEC:prediction:error} at the decoder coincide with those at the encoder, \eqref{eq:scheme:ENC:estimate:estimator} and \eqref{eq:scheme:ENC:filters:P}, respectively, 
    recovering the renowned results of classical (without a communication channel) LQG control.
\end{remark}

We next discuss the special cases of $\Kc = 1$ and $\Kc = 2$ channel uses per sample in Sections \ref{ss:LQG:Goblick} and \ref{ss:LQG:JSCC}, respectively.


\subsection{Source--Channel Rate Match}
\label{ss:LQG:Goblick}

\begin{figure}[t]
\centering
    \includegraphics[width=\columnwidth]{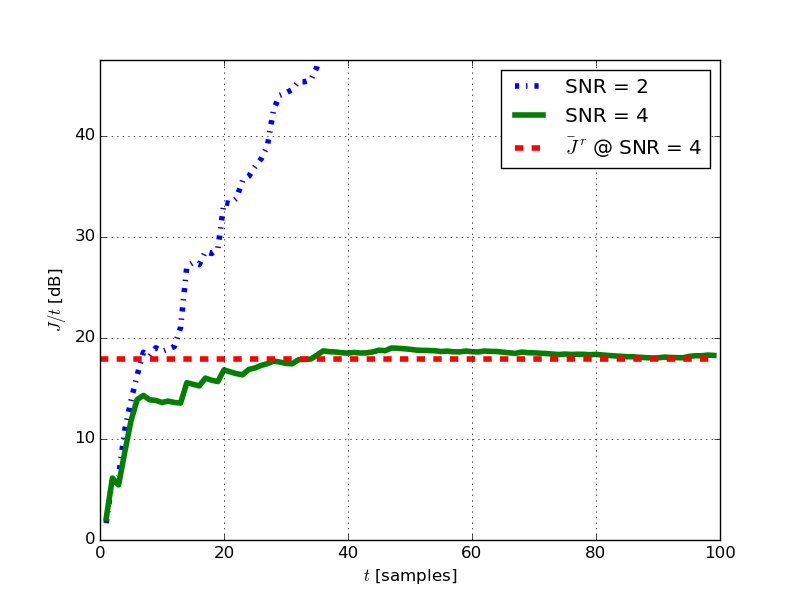}
\centering
    \caption{Optimal average stage LQG cost $\Jdec$ of a single representative run for $\Kc = \Ks = 1$, $\alpha = 2$, and SNRs 2 and 4 which 
    correspond to a stabilizable and an unstabilizable systems.
    The driving noise and observation noise powers and the LQG penalty coefficients are $Q=R=F=W=V=1$.}
    \label{fig:LQG:Goblick}
\end{figure}

In this subsection we treat the case of $\Kc = 1$, namely, where the sample rate of the control system and the signaling rate of the communication channel match.

As we saw in \secref{s:JSCC}, analog linear transmission of a Gaussian source over an AWGN channel 
achieves optimal performance (even when infinite delay is allowed), namely, the OPTA SDR~\eqref{eq:SDR:OPTA}, and given any input value.
Thus, the JSCC scheme that we use in this case is linear transmission~--- the source is transmitted as is, up to a power adjustment [recall~\eqref{eq:LQG:source} and~\eqref{eq:LQG:source:normalized}]:
\begin{align}
    a_t &= \us_t 
 \\ &= \frac{1}{\sqrt{\Pdec_{t|t-1} - \Penc_{t|t}}} s_t .
\end{align}
Since in this case $\SDR_0 = \SDRopta$, the upper and lower bounds of Theorems \ref{thm:UB} and \ref{thm:LB} coincide, establishing the optimum performance in  this case.
\vspace{.2\baselineskip}
\begin{corol}
\label{corol:rate_match}
    The scalar control system of \secref{s:model} with $\Kc = \Ks = 1$
    is stabilizable if only if $\alpha^2 < 1 + \SNR$, 
    and the optimal achievable infinite-horizon average stage LQG cost satisfies \eqref{eq:thm:UB:cost:DEC} with equality where $\SDR_0 = \SNR$.
\end{corol}
\vspace{.3\baselineskip}

\begin{remark}
    The stabilizability condition and optimum MMSE performance were previously established in \cite{LQGoverAWGN:linear:WithBraslavsky,LQGoverAWGN:linear:NoBraslavsky}
    for the case of no observation noise $V=0$. 
\end{remark}
    
\vspace{.5\baselineskip}
The optimal averate stage LQG cost
is illustrated in \figref{fig:LQG:Goblick}, where the normalized in time LQG cost $\Jdec$ is evaluated for a system with $\alpha = 2$ and two SNRs~--- 2 and 4. $\SNR = 4$ satisfies the stabilizability condition $\alpha^2 < 1 + \SNR$, 
whereas $\SNR = 2$ fails to do so. Unit LQG penalty coefficients \mbox{$Q=R=F=1$} and unit driving noise and observation noise powers $W=V=1$ are used.


\subsection{Source--Channel Rate Mismatch}
\label{ss:LQG:JSCC}

We now consider the case of $\Kc = 2$ channel uses per sample.
As we saw in \secref{s:JSCC}, linear schemes are suboptimal outside the low-SNR region.
Instead, by using non-linear maps, \eg, the (modified) Arcimedean spiral-based SK maps~\eqref{eq:spiral:bounded}, better performance can be achieved.

We note that the improvement in the SDR of the JSCC scheme is substantial when $\alpha^2$ is of the order of 
$\SDR$. That is, when the $\SDR$ of the linear scheme is close to $\alpha^2-1$, 
using an improved scheme with better $\SDR$ improves substantially the LQG cost.
Unfortunately, the spiral-based SK schemes do not promise any improvement for SNRs below 5dB under maximum-likelihood (ML) decoding.

\begin{remark}
    By replacing the ML decoder with an MMSE one, strictly better performance can be achieved over the linear scheme for all SNR values.
\end{remark}

\begin{figure}[t]
\centering
    \includegraphics[width=\columnwidth]{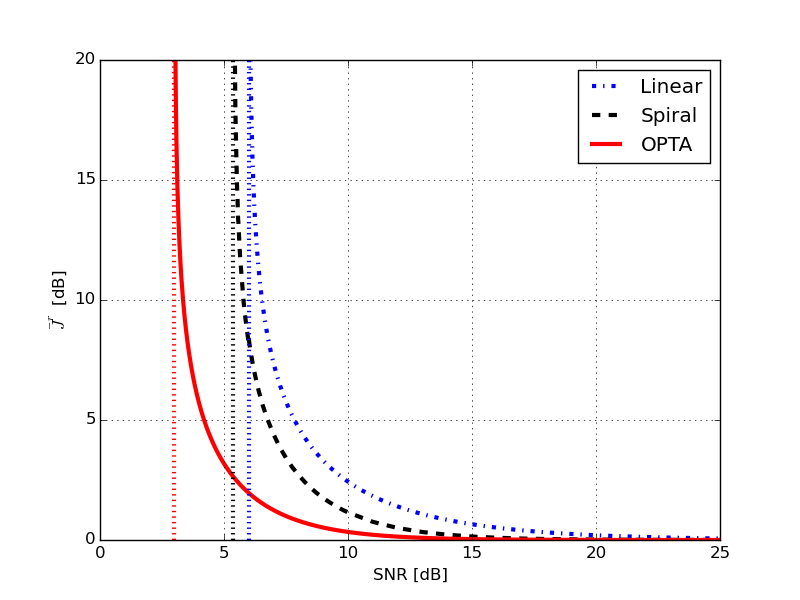}
\centering
    \caption{Average stage LQG costs when using the (distortion-bounded) SK Archimedean bi-spiral, repetition and the lower bound of~\thmref{thm:LB} for 
    \mbox{$\alpha = 3, W = 1, V = 0, Q = 1, R = 0$}. The vertical dotted lines represent the minimum SNR
    below which the cost diverges to infinity.}
    \label{fig:LQG:k=2}
\end{figure}

The effect of the $\SDR$ improvement is illustrated in \figref{fig:LQG:k=2} 
for a fully-observable ($V = 0$) system with $\alpha = 3$ and \mbox{$W = 1$}, for $Q = 1$ and $R = 0$, 
by comparing the achievable costs and lower bound of 
Theorems~\ref{thm:UB} and~\ref{thm:LB}.

\begin{remark}
    The resulting effective noise at the output of the JSCC receiver is not necessarily Gaussian, 
    and hence the resulting system states $x_t$, are not necessarily Gaussian either.
    Nevertheless, for the bounded-distortion scheme~\eqref{eq:spiral:bounded}, this has no effect on the resulting performance.
\end{remark}

%% file: summary.tex
In this paper we considered the simplest case of scalar systems, and $\Ks = 1$ and $\Kc = 2$.
Clearly, an (exponentially) large gain in performance can be achieved for $\Kc > 2$.

We further note that the results of Theorems~\ref{thm:UB} and~\ref{thm:LB} readily extend 
to systems with \emph{vector} states $\bx_t$ and \emph{vector} control signals $\bu_t$ but \emph{scalar} observed outputs $y_t$.

Interestingly, for the case of vector observed, state and control signals, 
even if the signaling rate of the channel and the sample rate of the observer are equal (rate matched case), 
conveying several analog observations over a single channel input may be of the essence. 
This is achieved by a compression JSCC scheme, \eg, 
by reversing the roles of the source and the channel inputs in the SK spiral-based scheme and similarly promises exponentially growing gains with the SNR and dimension; 
see
\cite{Shannon49,KotelnikovJSCC,ShannonKotelnikovMaps_Ramstad,ShannonKotelnikovMaps_MMSE_Dec,ChungPhD,AkyolViswanathaRoseRamstad_IterativeJSCC,IngberFederJSCC_DCC}.

In this work, we assumed that the observer knows all past control signals.
This case can be viewed as a two-sided side-information scenario. 
Nevertheless, although this is a common situation in practice, 
there are scenarios in which the observer is oblivious of the control signal applied or has only a noisy measurement of control signal generated by the controller. 
Such settings can be regarded as a JSCC problem with side information at the receiver (only), 
and can be treated using JSCC techniques designed for this case, some of which combine naturally with the JSCC schemes for rate mismatch~\cite{JointWZ-WDP,Tuncel_ZeroDelayJSCCwithWZ,AkyolViswanathaRoseRamstad_IterativeJSCC}.
In fact, this idea was recently applied for the related problem of communication over an AWGN channel with AWGN feedback in~\cite{BenYishaiShayevitz:FeedbackWithWZ:Allerton}

Finally, note that for the case of bounded noise (even worst-case/arbitrary bounded noise), using Hilbert curves (see, \eg, \cite{ChungPhD}) can provide a desirable solution and is extendable for any $1 : \Kc$ and $\Ks : 1$ rate ratios.